\documentclass[journal=jpcafh,manuscript=article]{achemso}
\usepackage[version=3]{mhchem} % Formula subscripts using \ce{}
\usepackage{braket}
\usepackage{amsmath}
\usepackage{color}
\usepackage{graphicx}% Include figure files

\author{Kenichi L. Ishikawa}
\email{ishiken@n.t.u-tokyo.ac.jp}
\affiliation[NEM]
{Department of Nuclear Engineering and Management, Graduate School of Engineering, The University of Tokyo,7-3-1 Hongo, Bunkyo-ku, Tokyo 113-8656, Japan}
\alsoaffiliation[PSC]
{Photon Science Center, Graduate School of Engineering, The University of Tokyo, 7-3-1 Hongo, Bunkyo-ku, Tokyo 113-8656, Japan}
\alsoaffiliation[UTripl]
{Research Institute for Photon Science and Laser Technology, The University of Tokyo, 7-3-1 Hongo, Bunkyo-ku, Tokyo 113-0033, Japan}
\alsoaffiliation[iALFA]
{Institute for Attosecond Laser Facility, The University of Tokyo, 7-3-1 Hongo, Bunkyo-ku, Tokyo 113-0033, Japan}
\author{Kevin C. Prince}
\affiliation[Elettra]
{Elettra-Sincrotrone Trieste, Basovizza 34149, Italy}
\author{Kiyoshi Ueda}
\affiliation[Tohoku University]
{Tohoku University, Sendai 980-8577, Japan}
\title[An \textsf{achemso} demo]
  {Control of ion-photoelectron entanglement and coherence via Rabi oscillations}
\abbreviations{IR,NMR,UV}

\begin{document}

% \begin{tocentry}

% Some journals require a graphical entry for the Table of Contents.
% This should be laid out ``print ready'' so that the sizing of the

% Inside the \texttt{tocentry} environment, the font used is Helvetica
% 8\,pt, as required by \emph{Journal of the American Chemical

% The surrounding frame is 9\,cm by 3.5\,cm, which is the maximum
% permitted for  \emph{Journal of the American Chemical Society}
% graphical table of content entries. The box will not resize if the
% content is too big: instead it will overflow the edge of the box.

% This box and the associated title will always be printed on a
% separate page at the end of the document.

% \end{tocentry}

\begin{abstract}
We report a theoretical investigation of photoionization by a pair of coherent, ultrashort fundamental and second-harmonic extreme-ultraviolet pulses, where the photon energies are selected to yield the same photoelectron energy for ionization of two different sub-shells. 
This choice implies that the fundamental energy is equal to the difference in energy of the ionic states, and they are therefore coupled by the fundamental photon. 
By deriving analytical expressions using the essential-states approach, we show that this Rabi coupling creates coherence between the two photoelectron wave packets, which would otherwise be incoherent. We analyze how the coupling is affected by the parameters such as relative phase, pulse width, delay between the two pulses, the Rabi coupling strength, and photoelectron energy. 
Our discussion mostly considers Ne $2p$ and $2s$ photoionization, but it is generally valid for many other quantum systems where photoionization from two different shells is observed.
\end{abstract}

\section{Introduction}
\label{sec:introduction}

Ultrashort (femtosecond and attosecond) extreme-ultraviolet (XUV) and soft x-ray (SX) pulses produced by free-electron lasers (FEL) \cite{Ackermann2007NPhoton,Allaria2012NPhoton,Allaria2014PRX,   Callegari2021PR} and high-harmonic generation (HHG) \cite{Ferray1988JPB,McPherson1987JOSAB,Paul2001Science,Hentschel2001Nature,Sekikawa2004Nature,Takahashi2013NComm,Fu2020CP,Midorikawa2022NPhoton} enable us to investigate a wide range of phenomena with chemical, physical and biological applications.
With longitudinally or temporally coherent pulses of two or more different wavelengths, e.g., available from the free-electron laser FERMI \cite{Allaria2012NPhoton,Allaria2014PRX,Prince2016NPhoton,Callegari2021PR,Perosa2023PRL}, scientists can coherently control the outcome of experiments by varying their relative phase or delay \cite{Prince2016NPhoton,You2016PRA,Giannessi2018,DiFraia2019PRL,You2020PRX}.

For example, Ne $2p$ ionization by two fundamental photons ($\omega + \omega$ scheme) and a single second-harmonic photon ($2\omega$ scheme) emits photoelectrons with the same energy but with opposite parities.
Since the two ionization pathways interfere with each other, the photoelectron angular distribution (PAD) and its asymmetry oscillate as a function of the $\omega$-$2\omega$ relative phase. 
This coherent control has been experimentally demonstrated \cite{Prince2016NPhoton} and applied to measurement of angle-resolved phases in photoemission \cite{You2020PRX}.

\begin{figure}
    \centering
    \includegraphics[width=0.8\hsize]{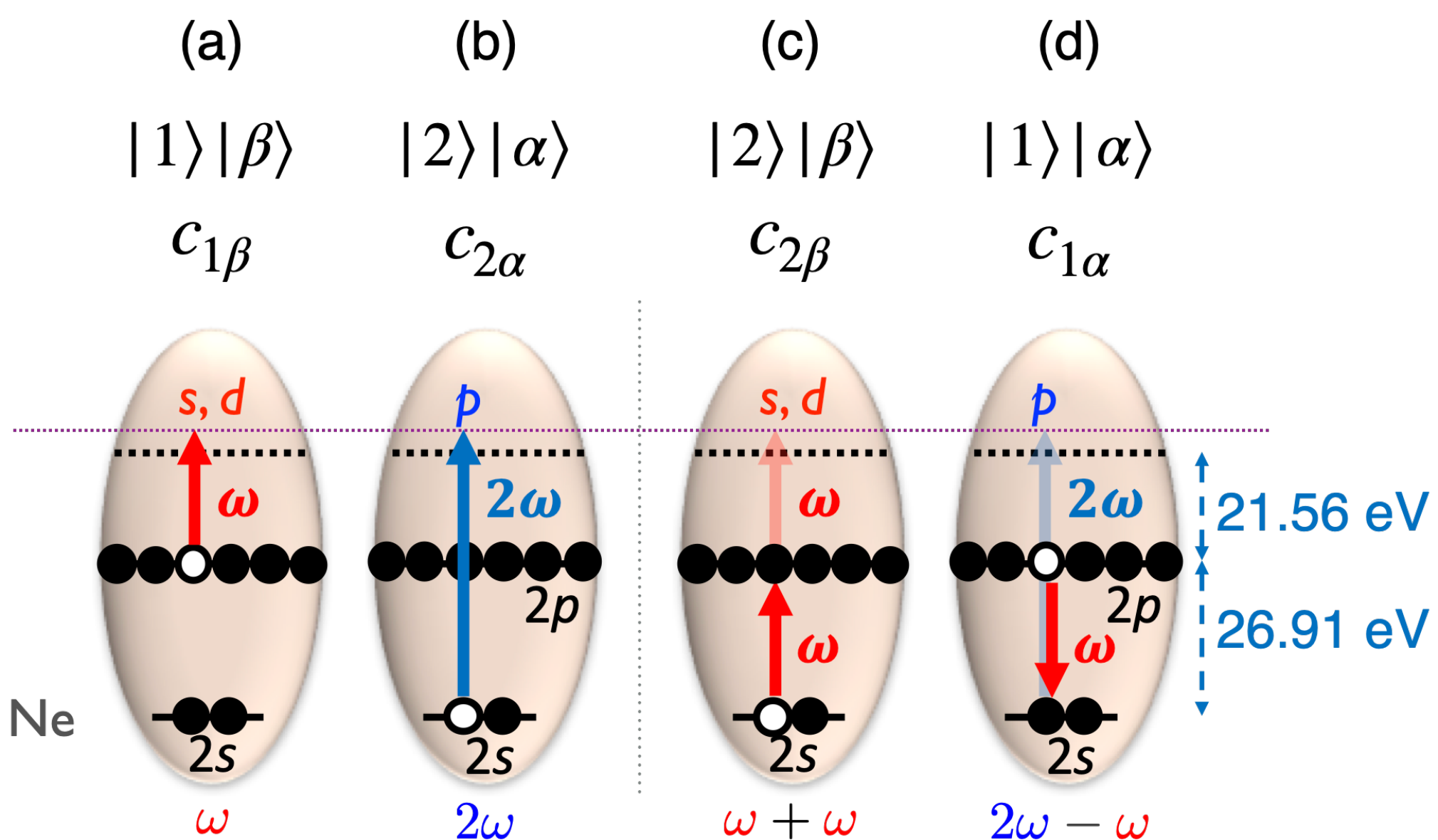}
    \caption{Schematic diagrams of processes considered in this study. (a) ionization of $2p$ by a $\hbar\omega$ photon (b) ionization of $2s$ by a $2\hbar\omega$ photon (c) excitation of $2s$ to $2p$ by a $\hbar\omega$ photon (d) de-excitation of $2p$ to $2s$ induced by a $\hbar\omega$ photon.
    $\ket{1}$ and $\ket{2}$ denote the ionic states $2p^{-1}$ and $2s^{-1}$, respectively, and $\ket{\alpha}$ and $\ket{\beta}$ denote the photoelectrons ejected from $2s$ ($p$ angular momentum) and $2p$ ($s$ and $d$ angular momenta).}
    \label{fig:scheme}
\end{figure}

Let us now consider the specific case of a Ne atom subject to collinearly $z$-polarized bichromatic light of the fundamental plus second-harmonic wavelength, i.e., $\hbar\omega$ and $2\hbar\omega$.
The thresholds for Ne $2p$ and $2s$ ionization are 21.56 and 48.47 eV, respectively.
Thus, if $\hbar\omega$ is tuned to their difference (26.91 eV), (single-photon) ionization from $2s$ by $2\hbar\omega$ and that from $2p$ by $\hbar\omega$ results in the same photoelectron energy (5.35 eV) and opposite parities [Fig.~\ref{fig:scheme}(a) and (b)]. 
Let $\ket{\alpha}$ ($p$ angular momentum) and $\ket{\beta}$ ($s$ and $d$ angular momenta) denote each photoelectron state.
In this case, however, the photoelectron wave packets from the two pathways do not interfere with each other, since the photoelectron and ionic core constitute one composite whole quantum system, and the two ionization pathways leave different ionic states ($2s^{-1}$ and $2p^{-1}$, denoted by $\ket{2}$ and $\ket{1}$, respectively), or, in other words, the photoelectron and ionic subsystems are entangled as $c_{2\alpha}\ket{2}\ket{\alpha}+c_{1\beta}\ket{1}\ket{\beta}$, with $c_{2\alpha}$ and $c_{1\beta}$ begin the amplitudes of the corresponding states.
Such correlation or entanglement between the photoelectron and ionic states is attracting increasing attention in attosecond science \cite{Vrakking2021PRL,Vrakking2022JPB,Nabekawa2023PRR}.
Now, since $\hbar\omega$ is equal to the energy difference between the $2s^{-1}$ and $2p^{-1}$ states, the fundamental pulse couples the two ionic states, inducing Rabi oscillations \cite{Rabi1937PR,Flogel2017PRA,Yu2018PRA,Nandi2022Nature}.
If the $2p^{-1}$ ionic state ($\ket{1}$) is excited to $2s^{-1}$ ($\ket{2}$) (i.e., if the $2s$ electron is excited to $2p$) [Fig.~\ref{fig:scheme}(c)] or vice versa [Fig.~\ref{fig:scheme}(d)], the two photoelectron states $\ket{\alpha}$ and $\ket{\beta}$ can interfere with each other.
Hence, a question arises, ``does the Rabi coupling convert the entanglement to coherent superposition, leading to, e.g., coherent control of the photoelectron angular distribution through the $\omega$-$2\omega$ relative phase?"

In the present paper, we investigate photoionization of an atom by ultrashort fundamental and second-harmonic XUV pulses with the same photoelectron energy but two different ionic states, coupled by the fundamental photon.
Using an essential-states approach \cite{Eberly1991PR,Grobe1993PRA,Yu2018PRA}, we show that the Rabi coupling indeed creates coherence between the two, otherwise incoherent, photoelectron wave packets. Furthermore, we analyze how it is affected by different parameters such as the relative phase, pulse width, delay between the two wavelength components, the Rabi coupling strength, and photoelectron energy.
Whereas our discussion mostly takes Ne $2p$ and $2s$ photoionization as an example for the sake of concreteness, it will be generally valid for many other quantum systems where photoionization from two different shells is observed.

This paper is organized as follows. The second section describes our model based on the essential-states approach and derives the formula for the degree of coherence between the two ionization pathways. In the third section, we show numerical results and discuss how the degree of coherence depends on various parameters. Conclusions are given in the fourth section.
Atomic units (a.u.) are used throughout unless otherwise stated.

% This is a paragraph of text to fill the introduction of the
% demonstration file.  The demonstration file attempts to show the
% modifications of the standard \LaTeX\ macros that are implemented by
% the \textsf{achemso} class.  These are mainly concerned with content,
% as opposed to appearance.

\section{Theory}
\label{sec:essential-states}

\subsection{Essential-states approach}

Let us consider the ground-state neutral atom $\ket{g}$ and the four states and processes schematically depicted in Fig.~\ref{fig:scheme}.
Since we are interested in the interference of photoelectron wave packets that can be coherently controlled by the $\omega$-$2\omega$ relative phase $\phi$, we focus on the processes involving the photoelectrons with the same spin, magnetic angular momentum $m=0$, and energy centered at 5.35 eV.
Non-interfering contributions from ionization of $2p(m=\pm 1)$ electrons by a $\hbar\omega$ photon and that of $2p$ electrons by a $2\hbar\omega$ photon can be taken into account by including the ground-state depletion and also in the calculation of the photoelectron angular distribution.
Whereas these contributions are neglected in this section for simplicity, they will be included in numerical examples for Ne presented further below (see Eqs.~\ref{eq:PAD}, \ref{eq:PAD2}, \ref{eq:beta2}, and \ref{eq:beta4}).

Following Refs. \cite{Grobe1993PRA,Yu2018PRA}, 
% https://journals.aps.org/pra/abstract/10.1103/PhysRevA.48.623
% https://doi.org/10.1103/PhysRevA.98.033404
we expand the time-dependent two-electron wave function $\ket{\Psi (t)}$ as a linear combination of the essential states, i.e., the ground state $\ket{g}$ and four singly ionized continua $\ket{1\,\alpha;\epsilon}, \ket{2\,\alpha;\epsilon}, \ket{1\,\beta;\epsilon}, \ket{2\,\beta;\epsilon}$ (for example, $\ket{1\,\alpha;\epsilon}$ denotes a singly ionized continuum state composed of the ionic state $\ket{1}$ [$2p^{-1}$] and the photoelectron state $\ket{\alpha}$ [$p$ angular momentum] with an energy of $\epsilon$):
\begin{align}
    \ket{\Psi(t)} = c_g(t) e^{-i\omega_g t}\ket{g} &+ \int c_{1\beta}(\Delta\epsilon,t) e^{-i(\omega_g+\omega+\Delta\epsilon)t}\ket{1\,\beta;\epsilon}d\epsilon \nonumber\\
    &+ \int c_{2\beta}(\Delta\epsilon,t) e^{-i(\omega_g+2\omega+\Delta\epsilon)t}\ket{2\,\beta;\epsilon}d\epsilon \nonumber\\
    &+ \int c_{2\alpha}(\Delta\epsilon,t) e^{-i(\omega_g+2\omega+\Delta\epsilon)t}\ket{2\,\alpha;\epsilon}d\epsilon \nonumber\\
    &+ \int c_{1\alpha}(\Delta\epsilon,t) e^{-i(\omega_g+\omega+\Delta\epsilon)t}\ket{1\,\alpha;\epsilon}d\epsilon \label{eq:essential-state-expansion},
\end{align}
where $\Delta\epsilon=\epsilon-(\hbar\omega-I_p)$ denotes the difference of the photoelectron energy $\epsilon$ from its central value $\hbar\omega-I_p$, with $I_p$ being the first ionization potential.
The bichromatic electric field $E(t)$ is described by,
\begin{equation}
    E(t) = F_\omega (t) \cos\omega t + F_{2\omega} (t) \cos (2\omega t - \phi),
    \label{eq:pulse-shape}
\end{equation}
where $F_\omega (t)$ and $F_{2\omega} (t)$ denote the pulse envelopes, and $\phi$ the $\omega$-$2\omega$ relative phase.
We assume the electric dipole approximation and employ the length gauge for the laser-electron interaction $H_I=(z_1+z_2)E(t)$.

We introduce the notations for the dipole couplings between $\ket{g}$ and continuum state $\ket{q;\epsilon}$,
\begin{equation}
    \braket{g|z_1+z_2|q;\epsilon} = D(g,q;\epsilon),
\end{equation}
and for the coupling between the two ionized states, approximated as \cite{Hanson1997PRA},
\begin{equation}
    \braket{1\,\alpha;\epsilon|z_1+z_2|2\,\alpha;\epsilon^\prime} = \braket{1\,\beta;\epsilon|z_1+z_2|2\,\beta;\epsilon^\prime} \approx D(1,2) \delta (\epsilon - \epsilon^\prime),
\end{equation}
with $D(1,2)$ being the dipole transition matrix element between the two ionic states, assumed to be real valued and positive.
If we substitute Eq.~\ref{eq:essential-state-expansion} into the time-dependent Schr\"odinger equation and project on each state, we obtain the following equations of motion (EOMs) for the coefficients that govern the temporal evolution of the system within the rotating-wave approximation (RWA): 
\begin{align}
    i\dot{c}_g(t) &= \frac{1}{2}\int D(g,1\,\beta;\epsilon)c_{1\beta}(\Delta\epsilon,t)F_\omega (t) e^{-i\Delta\epsilon t} d\epsilon \nonumber \\
    &+ \frac{1}{2}\int D(g,2\,\alpha;\epsilon)c_{2\alpha}(\Delta\epsilon,t)F_{2\omega} (t) e^{-i(\Delta\epsilon t + \phi)} d\epsilon \label{eq:EOM-i} \\
%    &+ \frac{1}{2}\int D(g,1\,\gamma;\epsilon) b(\Delta\epsilon,t) F_{2\omega} (t) e^{-i(\Delta\epsilon t + \phi)} d\epsilon,  \\
    i\dot{c}_{1\beta}(\Delta\epsilon,t) &= \frac{1}{2}\Omega(t) c_{2\beta}(\Delta\epsilon,t) + \frac{1}{2}D(g,1\,\beta;\epsilon)^* c_g(t)F_\omega (t) e^{i\Delta\epsilon t} \label{eq:EOM-1beta} \\
    i\dot{c}_{2\beta}(\Delta\epsilon,t) &= \frac{1}{2}\Omega(t)c_{1\beta}(\Delta\epsilon,t) \label{eq:EOM-2beta} \\
    i\dot{c}_{2\alpha}(\Delta\epsilon,t) &= \frac{1}{2}\Omega(t) c_{1\alpha}(\Delta\epsilon,t) + \frac{1}{2}D(g,2\,\alpha;\epsilon)^* c_g(t)F_{2\omega} (t) e^{i(\Delta\epsilon t + \phi)} \label{eq:EOM-2alpha} \\
    i\dot{c}_{1\alpha}(\Delta\epsilon,t) &= \frac{1}{2}\Omega(t) c_{2\alpha}(\Delta\epsilon,t) 
    \label{eq:EOM-1alpha}
\end{align}
with,
\begin{equation}
    \Omega (t) = D(1,2)F_\omega (t),
\end{equation}
being the time-varying Rabi frequency.
The terms containing $\Omega (t)$ in Eqs.~\ref{eq:EOM-1beta}, \ref{eq:EOM-2beta}, \ref{eq:EOM-2alpha}, and \ref{eq:EOM-1alpha} correspond to the Rabi oscillations.
% Ionization to $\ket{1\,\gamma;\epsilon}$ by $2\omega$ expressed by the last term on the right-hand side of Eq.~\ref{eq:EOM-i} is treated as a loss channel, since we are interested in the interference of different ionization channels with the common photoelectron energy $\epsilon \sim E_2+\hbar\omega$. 
% ($E_2$ is not defined - the definition may be in text that has been commented out. should it be or $\epsilon \sim \hbar\omega-I_p$)

By solving the EOMs, we obtain the following expressions for the amplitudes of the individual states,
\begin{align}
%     C_{1}(\Delta\epsilon, t\to\infty) &= -i\frac{\sqrt{2}}{4}D(g,2\,\alpha;\epsilon)^* e^{i\phi}\int_{-\infty}^\infty c_g(t)F_{2\omega}(t)\exp \left[i\left(\Delta\epsilon\,t-\frac{1}{2}\int_t^\infty \Omega (t^\prime) dt^\prime\right)\right] dt \\
% %
%     C_{1}(\Delta\epsilon, t) &= -i\frac{\sqrt{2}}{4}D(g,2\,\alpha;\epsilon)^* e^{i\phi}\int_{-\infty}^t c_g(t^\prime)F_{2\omega}(t^\prime)\exp \left[i\left(\Delta\epsilon\,t^\prime-\frac{1}{2}\int_{t^\prime}^t \Omega (t^{\prime\prime}) dt^{\prime\prime}\right)\right] dt^\prime \\
% %
%     C_{2}(\Delta\epsilon, t) &= -i\frac{\sqrt{2}}{4}D(g,2\,\alpha;\epsilon)^* e^{i\phi}\int_{-\infty}^t c_g(t^\prime)F_{2\omega}(t^\prime)\exp \left[i\left(\Delta\epsilon\,t^\prime+\frac{1}{2}\int_{t^\prime}^t \Omega (t^{\prime\prime}) dt^{\prime\prime}\right)\right] dt^\prime \\
%
    c_g(t) &= \exp\left[-\frac{\pi D(g,1\,\beta)^2}{4}\int_{-\infty}^t F_\omega^2 (t^\prime)^2 dt^\prime
    -\frac{\pi D(g,2\,\alpha)^2}{4}\int_{-\infty}^t F_{2\omega}^2 (t^\prime)^2 dt^\prime \right] \\
    % & \left. -\, {\rm (loss\,channel)}
    % \right] \\
%    
    c_{2\alpha}(\Delta\epsilon, t) &= -\frac{i}{2}D(g,2\,\alpha;\epsilon)^* e^{i\phi}\int_{-\infty}^t c_g(t^\prime)F_{2\omega}(t^\prime) \cos \left(\frac{1}{2}\int_{t^\prime}^t \Omega (t^{\prime\prime}) dt^{\prime\prime}\right) e^{i\Delta\epsilon\,t^\prime} dt^\prime \\
    c_{1\alpha}(\Delta\epsilon, t) &= -\frac{1}{2}D(g,2\,\alpha;\epsilon)^* e^{i\phi}\int_{-\infty}^t c_g(t^\prime)F_{2\omega}(t^\prime) \sin \left(\frac{1}{2}\int_{t^\prime}^t \Omega (t^{\prime\prime}) dt^{\prime\prime}\right) e^{i\Delta\epsilon\,t^\prime} dt^\prime \\
    c_{2\beta}(\Delta\epsilon, t) &= -\frac{1}{2}D(g,1\,\beta;\epsilon)^* \int_{-\infty}^t c_g(t^\prime)F_\omega(t^\prime)  \sin \left(\frac{1}{2}\int_{t^\prime}^t \Omega (t^{\prime\prime}) dt^{\prime\prime}\right) e^{i\Delta\epsilon\,t^\prime} dt^\prime \\
    c_{1\beta}(\Delta\epsilon, t) &= -\frac{i}{2}D(g,1\,\beta;\epsilon)^* \int_{-\infty}^t c_g(t^\prime)F_\omega(t^\prime)  \cos \left(\frac{1}{2}\int_{t^\prime}^t \Omega (t^{\prime\prime}) dt^{\prime\prime}\right) e^{i\Delta\epsilon\,t^\prime} dt^\prime
\end{align}
% with,
% \begin{equation}
%     \Omega (t) = D(1,2)F_\omega (t).
% \end{equation}
They become after the pulse ($t\to\infty$),
\begin{align}
    c_g(\Delta\epsilon, t\to\infty) &= \exp\left[-\frac{\pi D(g,1\,\beta)^2}{4}\int_{-\infty}^\infty F_\omega^2 (t)^2 dt
    -\frac{\pi D(g,2\,\alpha)^2}{4}\int_{-\infty}^\infty F_{2\omega}^2 (t)^2 dt \right] \\
    % & \left. -\, {\rm (loss\,channel)}
    % \right] \\
%    
    c_{2\alpha}(\Delta\epsilon, t\to\infty) &= -\frac{i}{2}D(g,2\,\alpha;\epsilon)^* e^{i\phi}\int_{-\infty}^\infty c_g(t)F_{2\omega}(t)\cos \left(\frac{1}{2}\int_{t}^\infty \Omega (t^\prime) dt^\prime\right) e^{i\Delta\epsilon\,t} dt \label{eq:c2alpha-final}\\
    c_{1\alpha}(\Delta\epsilon, t\to\infty) &= -\frac{1}{2}D(g,2\,\alpha;\epsilon)^* e^{i\phi}\int_{-\infty}^\infty c_g(t)F_{2\omega}(t) \sin \left(\frac{1}{2}\int_{t}^\infty \Omega (t^\prime) dt^\prime\right) e^{i\Delta\epsilon\,t} dt \label{eq:c1alpha-final}\\
    c_{2\beta}(\Delta\epsilon, t\to\infty) &= -\frac{1}{2}D(g,1\,\beta;\epsilon)^* \int_{-\infty}^\infty c_g(t)F_\omega(t) \sin \left(\frac{1}{2}\int_{t}^\infty \Omega (t^\prime) dt^\prime\right)  e^{i\Delta\epsilon\,t} dt \label{eq:c2beta-final}\\
    c_{1\beta}(\Delta\epsilon, t\to\infty) &= -\frac{i}{2}D(g,1\,\beta;\epsilon)^* \int_{-\infty}^\infty c_g(t)F_\omega(t) \cos \left(\frac{1}{2}\int_{t}^\infty \Omega (t^\prime) dt^\prime\right)  e^{i\Delta\epsilon\,t} dt \label{eq:c1beta-final}
\end{align}
Note that $c_g(t)F_{2\omega}(t)$ and $c_g(t)F_\omega(t)$ in the integrands physically correspond to photoionization, $e^{i\Delta\epsilon\,t}$ the effect of detuning within the band width, and $\cos \left(\frac{1}{2}\int_{t}^\infty \Omega (t^\prime) dt^\prime\right)$ and $\sin \left(\frac{1}{2}\int_{t}^\infty \Omega (t^\prime) dt^\prime\right)$ the Rabi couplings.

\subsection{Interference between the ionization pathways}

Suppose that the photoelectron wave functions in the $\alpha$ and $\beta$ channels have angular parts $Y_\alpha(\theta,\varphi)$ and $Y_\beta(\theta,\varphi)$, respectively, with opposite parities.
Associated with the same ionic core, the interference between $\ket{1\,\alpha}$ and $\ket{1\,\beta}$ as well as that between $\ket{2\,\alpha}$ and $\ket{2\,\beta}$ can be coherently controlled through the $\omega$-$2\omega$ relative phase $\phi$ and detected by the photoelectron angular distribution (PAD). Hence, the interference is characterized by $c_{1\alpha}(\Delta\epsilon,\infty) \left[c_{1\beta}(\Delta\epsilon,\infty)\right]^*+c_{2\alpha}(\Delta\epsilon,\infty) \left[c_{2\beta}(\Delta\epsilon,\infty)\right]^*$.
Formulated more rigorously, if we take the trace of the density operator of the composite photoelectron-ion system with respect to the ionic state, by taking electron exchange into account, we obtain the photoelectron reduced density operator,
\begin{align}
\label{eq:reduced-density-operator}
   & \ket{\alpha}(|c_{1\alpha}|^2+|c_{2\alpha}|^2)\bra{\alpha}
    - \ket{\alpha}(c_{1\alpha}c_{1\beta}^* + c_{2\alpha}c_{2\beta}^*) \bra{\beta} \nonumber \\
   & - \ket{\beta}(c_{1\alpha}^*c_{1\beta} + c_{2\alpha}^* c_{2\beta}) \bra{\alpha}
    + \ket{\beta}(|c_{1\beta}|^2+|c_{2\beta}|^2)\bra{\beta}.
\end{align}
Therefore, the degree of coherence can by evaluated as,
\begin{equation}
\label{eq:degree-of-interference}
I_{\rm coh}(\Delta\epsilon)=-\frac{c_{1\alpha}c_{1\beta}^* + c_{2\alpha}c_{2\beta}^*}{|c_{1\alpha}|^2+|c_{2\alpha}|^2+|c_{1\beta}|^2+|c_{2\beta}|^2},
\end{equation}
which takes a complex value.
In Eqs.~\ref{eq:reduced-density-operator} and \ref{eq:degree-of-interference} the dependence of the channel amplitudes on $\Delta\epsilon$ is omitted for simplicity. 
If the Rabi coupling is negligible, i.e., $D(1,2) = 0$, $c_{1\alpha}$ and $c_{2\beta}$ vanish, and as a result, $I_{\rm coh}(\Delta\epsilon)$ also vanishes; the $\alpha$ and $\beta$ photoelectrons would not interfere with each other, since they are entangled with different ionic cores. 
$I_{\rm coh}$ varies with $\phi$ as,
\begin{equation}
    \label{eq:Icoh-phi-dependence}
    I_{\rm coh}(\Delta\epsilon) = I_{\rm coh}(\Delta\epsilon)|_{\phi=0} \, e^{i\phi}.
\end{equation}
As we will see below (Eqs.~\ref{eq:beta1} and \ref{eq:beta3}), $I_{\rm coh}$ indeed characterizes the PAD oscillation with $\phi$.

It is worth mentioning that the modulation by the Rabi couplings in Eqs.~\ref{eq:c2alpha-final} and \ref{eq:c1alpha-final}, which convert the entanglement to coherence, involve temporal integration to infinity. Thus, even if the $2\omega$ pulse precedes the $\omega$ pulse with no overlap, the photoelectron which was ejected by the former, even when distant from the parent ion, is influenced by the delayed $\omega$ pulse acting not directly on the photoelectron but on the remaining ion.
This takes place not due to the long-range Coulomb interaction but due to the quantum-mechanical integrity or nonlocal nature of the whole ion-photoelectron system.

\section{Results and discussion}
\label{sec:results}

\subsection{General features}

\begin{figure}
    \centering
    \includegraphics[width=0.8\hsize]{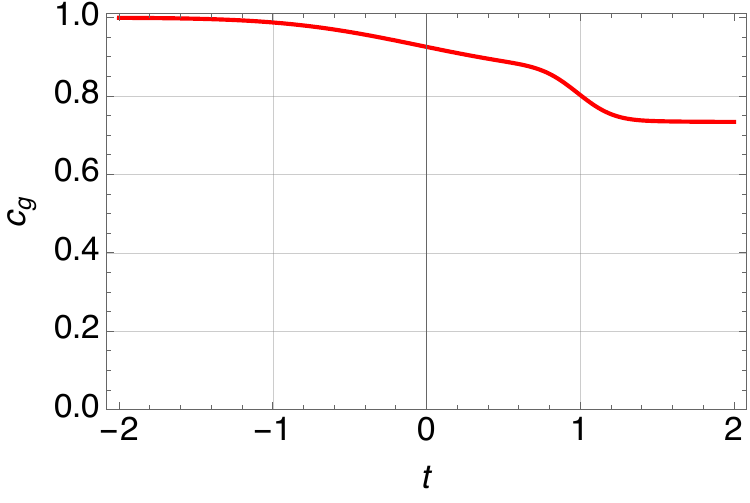}
    \caption{Time evolution of the initial state amplitude $c_g(t)$ for $T=\frac{1}{4}$ and $\tau=1$}
    \label{fig:ci-time-evolution}
\end{figure}

To get an insight into the general behavior of $I_{\rm coh}(\Delta\epsilon)$, let us first consider {\it normalized, dimensionless} model cases, in which the fundamental pulse envelope is expressed as,
\begin{equation}
    F_\omega (t) = \exp \left(-\frac{t^2}{2}\right).\label{eq:model-w-pulse}
\end{equation}
It should be noticed that atomic units are {\it not} used in this Sub-section. Instead, time and electric field are normalized in such a way that the temporal profile of the fundamental pulse is described by Eq.~\ref{eq:model-w-pulse}. Then, $\Delta\epsilon$ is normalized so that the full-width-at-half-maximum (FWHM) band width of the $\omega$ pulse is $2\sqrt{\ln 2}(=1.67)$, assumed to be sufficiently smaller than the level spacing. We further introduce,
\begin{align}
    F_{2\omega} (t) & = \frac{1}{\sqrt{T}}\exp \left[-\frac{(t-\tau)^2}{2T^2}\right], \\
    D(g,1\,\beta) &= D(g,2\,\alpha) = \frac{1}{3}, \\
    D(1,2) &= \sqrt{2\pi}\, \xi,
\end{align}
which model physical situations of comparable ionization rates from $2s$ and $2p$ and moderate ground-state depletion.
$\xi$ is a scaling factor, so that $\int_{-\infty}^\infty \Omega (t) dt = 2\pi\xi$, and  $T$ and $\tau$ denote the pulse width and delay of the second-harmonic pulse relative to the fundamental, respectively.
Both pulses have the same fluence $\int_{-\infty}^\infty F_\omega (t)^2 dt = \int_{-\infty}^\infty F_{2\omega} (t)^2 dt$.
Since ionization is of single-photon nature, the final value of $c_g(t)$ is $e^{-\pi^{3/2}/18}=0.734$, independent of $T$ and $\tau$, corresponding to a final ground-state population 0.539 after both pulses.
In Fig.~\ref{fig:ci-time-evolution}, we show an example of $c_g(t)$ for $T=\frac{1}{4}$ and $\tau=1$.

\subsubsection{Dependence on the Rabi coupling strength $\xi$}

\begin{figure}
    \centering
    \includegraphics[width=0.8\hsize]{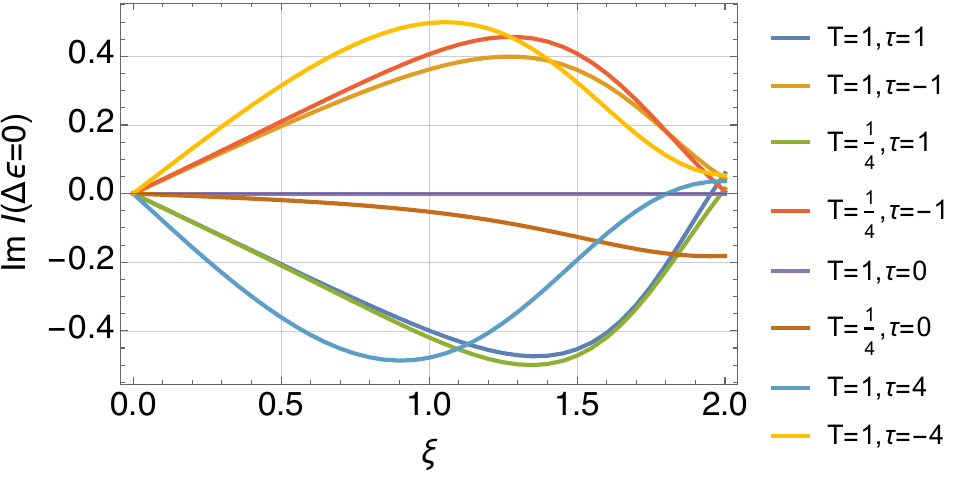}
    \caption{$\xi$-dependence of $\operatorname{Im}I_{\rm coh}(\Delta\epsilon=0)$ for $\phi=0$ and different combinations of $T$ and $\tau$.}
    \label{fig:xi-dependence}
\end{figure}

\begin{figure}
    \centering
    \includegraphics[width=0.8\hsize]{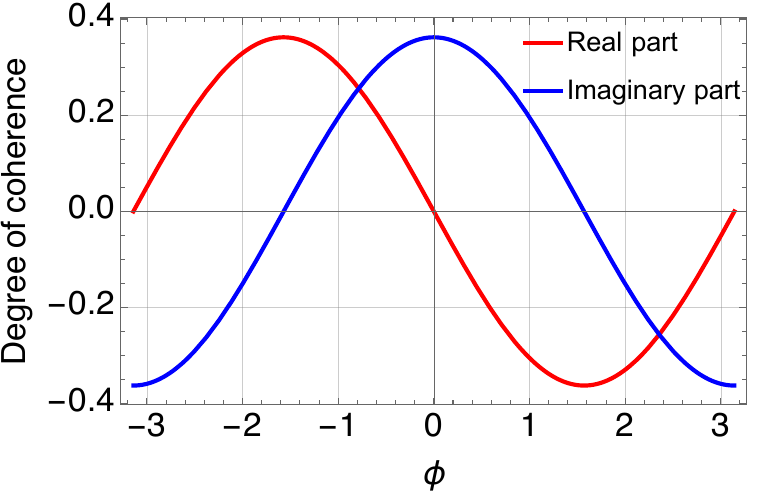}
%    Ne2s2p20210504-7bisbis.nb
    \caption{$\phi$-dependence of $\operatorname{Re}I_{\rm coh}(\Delta\epsilon=0)$ and $\operatorname{Im}I_{\rm coh}(\Delta\epsilon=0)$ for $\xi=1$, $T=1$ and $\tau=-1$.}
    \label{fig:phi-dependence}
\end{figure}

At $\Delta\epsilon=0$ (spectral center) and $\phi=0$, the complex degree of coherence is pure imaginary. Its dependence on $\xi$ for different combinations of $T$ and $\tau$ is shown in Fig.~\ref{fig:xi-dependence}.
At $\xi=0$, i.e, without Rabi oscillations, there would be no pathway interference, as discussed above.
At $\xi > 0$, except for $T=1$ and $\tau=0$, we indeed find that coherence is created by the Rabi oscillations in general, as we expected.
In such cases, the real and imaginary parts oscillate with the $\omega$-$2\omega$ relative phase $\phi$ (Fig.~\ref{fig:phi-dependence}), which can be, in principle, observed through oscillation of the energy-resolved PAD with $\phi$.

\subsubsection{Delay dependence}

\begin{figure}
    \centering
    \includegraphics[width=0.8\hsize]{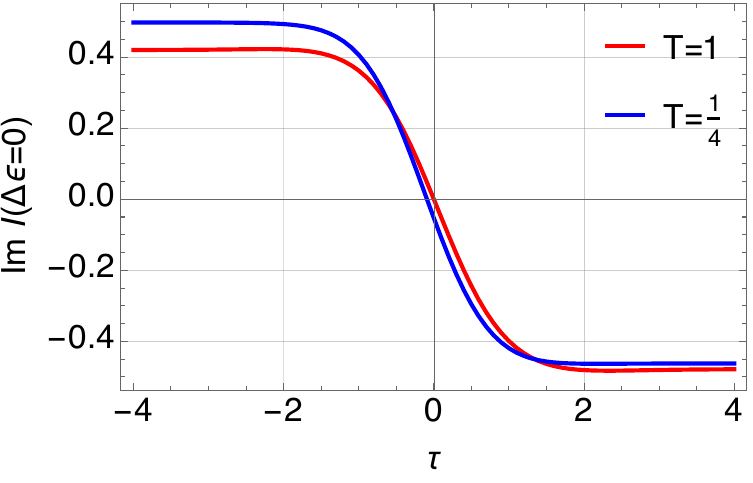}
    \caption{delay-dependence of $\operatorname{Im}I_{\rm coh}(\Delta\epsilon=0)$ for $\phi=0$, $\xi=1$, and $T=1$ and $\frac{1}{4}$. Note that the real part vanishes for $\Delta\epsilon=0$ and $\phi=0$.}
    \label{fig:delay-dependence}
\end{figure}

In Fig.~\ref{fig:xi-dependence}, at finite delay ($\tau=\pm 1$), the magnitude of the degree of coherence first increases with $\xi$, reaches its maximum between $\xi=1$ and 1.5, then decreases, and nearly vanishes at $\xi\approx 2$, suggesting one whole Rabi cycle on average.
On the other hand, when the two pulses overlap with each other ($\tau=0$), the degree of coherence is small, and in particular, completely vanishes if they have the same width ($T=1$). 
We analyze the delay dependence in Fig.~\ref{fig:delay-dependence}, where only the imaginary part is plotted, as the real part vanished for $\Delta\epsilon=0$ and $\phi=0$.
The degree of coherence passes through zero exactly at $\tau = 0$ for $T=1$ and near $\tau = 0$ for $T=\frac{1}{4}$, while its magnitude increases with increasing delay.
Using Eqs.~\ref{eq:c2alpha-final}--\ref{eq:c1beta-final} and \ref{eq:degree-of-interference}, one can easily show that $I_{\rm coh}(\Delta\epsilon=0)=0$, if the two pulses have similar forms, i.e., $F_{2\omega}(t) = k F_\omega (t)$ with a constant $k$, for any arbitrary relative phase $\phi$ and pulse shape, and is thus not limited to a Gaussian shape.
This is because, in such a case, the Rabi oscillation following $2p$ ionization and that following $2s$ ionization are represented by two points on the Bloch sphere that are always diametrically opposite.

\subsubsection{Photoelectron energy ($\Delta\epsilon$) dependence}

\begin{figure}
    \centering
    \includegraphics[width=0.8\hsize]{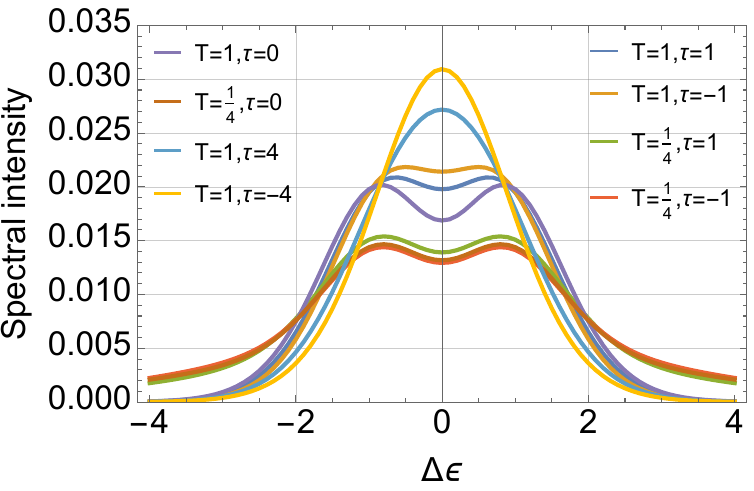}
    \caption{Angle-integrated photoelectron energy spectra for $\phi=0$, $\xi=1$, and different combinations of $T$ and $\tau$.}
    \label{fig:photoelectron-energy-spectrum}
\end{figure}

\begin{figure}
    \centering
    \includegraphics[width=0.8\hsize]{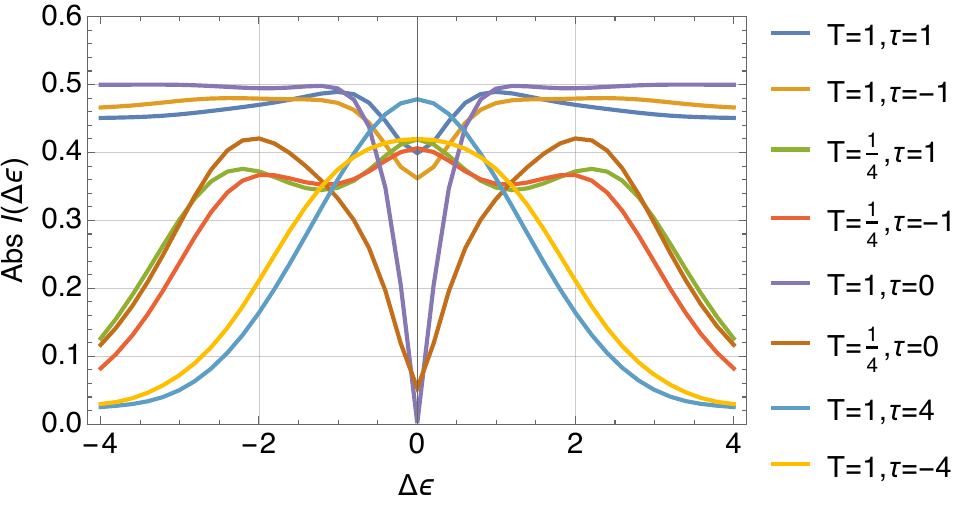}
    \caption{Photoelectron energy-dependence of the magnitude of the complex degree of coherence for $\phi=0$, $\xi=1$, and different combinations of $T$ and $\tau$.}
    \label{fig:energy-dependence-magnitude}
\end{figure}

\begin{figure}
    \centering
    \includegraphics[width=\hsize]{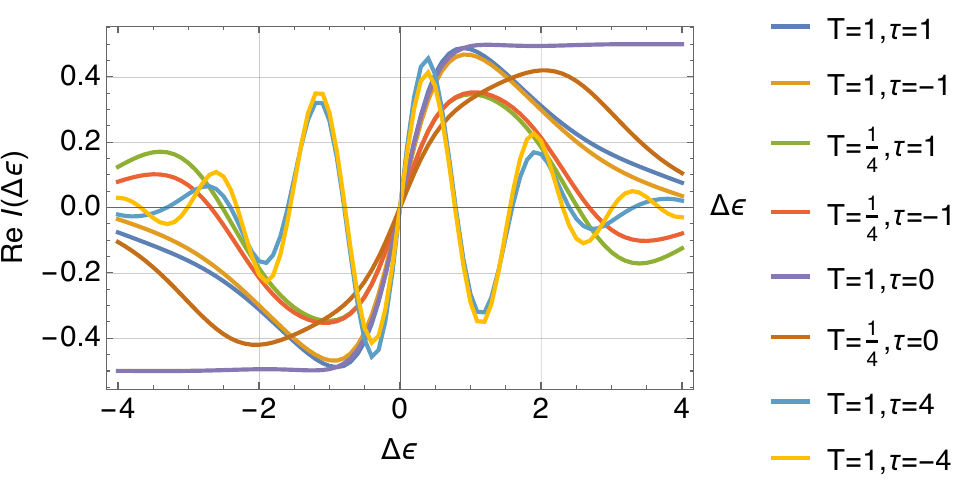}
    \caption{Photoelectron energy-dependence of the real part of the complex degree of coherence for $\phi=0$, $\xi=1$, and different combinations of $T$ and $\tau$.}
    \label{fig:energy-dependence-real}
\end{figure}

We have so far focused on the behavior of pathway interference at the spectral center $\Delta\epsilon=0$. Let us now turn to its dependence on $\Delta\epsilon$.
Figure \ref{fig:photoelectron-energy-spectrum} displays the photoelectron energy spectra $|c_{1\alpha}(\Delta\epsilon,\infty)|^2+|c_{2\alpha}(\Delta\epsilon,\infty)|^2+|c_{1\beta}(\Delta\epsilon,\infty)|^2+|c_{2\beta}(\Delta\epsilon,\infty)|^2$.
Note that the FWHM band width of the $\omega$ and $2\omega$ pulses is $2\sqrt{\ln 2}(=1.67)$ and  $2\sqrt{\ln 2}/T$, respectively, in this dimensionless model.
We can see spectral broadening and Autler-Townes \cite{Autler1955PR} peak splitting due to modulation by the Rabi oscillations, similar to what has previously been reported for single-color interaction with He \cite{Yu2018PRA,Nandi2022Nature}.
We plot the magnitude and real part of the complex degree of coherence in Figs.~\ref{fig:energy-dependence-magnitude} and \ref{fig:energy-dependence-real}.
Even for the no-delay cases, for which the degree of coherence vanishes for the spectral center, the pathway interference is revived for $\Delta\epsilon\ne 0$ (Fig.~\ref{fig:energy-dependence-magnitude}).
The real part $\operatorname{Re}I_{\rm coh}(\Delta\epsilon)$ oscillates with $\Delta\epsilon$ (Fig.~\ref{fig:energy-dependence-real}), reflecting the variation of $\arg I_{\rm coh}(\Delta\epsilon)$ with photoelectron energy.
In particular, when the two pulses are well separated (e.g., $\tau=4$), $\operatorname{Re}I_{\rm coh}(\Delta\epsilon)$ oscillates with a period of $\frac{2\pi}{\tau}$.

% \begin{figure}
%     \centering
%     \includegraphics[width=0.8\hsize]{figs/energy-dependence-arg.pdf}
%     \caption{Photoelectron energy-dependence of the argument of the complex degree of coherence for $\phi=0$, $\xi=1$, and different combinations of $T$ and $\tau$.}
%     \label{fig:energy-dependence-arg}
% \end{figure}

\subsubsection{Photoelectron-energy-integrated degree of coherence}

As we have seen in Fig.~\ref{fig:energy-dependence-real}, the real part of the degree of coherence, i.e., the phase of the PAD oscillation with $\phi$, oscillates with $\Delta\epsilon$.
Let us now investigate whether an oscillation with $\phi$ survives after integration of the PAD with respect to photoelectron energy $\epsilon$.
We can reasonably assume that the dipole couplings are independent of photoelectron energy $\epsilon$ within the spectral width of the pulse.
Then, Eqs.~\ref{eq:c2alpha-final}--\ref{eq:c1beta-final} can also be viewed as the Fourier transform of time-domain wave packets,
\begin{align}
    \hat{c}_{2\alpha}(t) &= -\frac{i}{2}D(g,2\,\alpha)^* e^{i\phi} c_g(t)F_{2\omega}(t)  \cos \left(\frac{1}{2}\int_{t}^\infty \Omega (t^\prime) dt^\prime\right) , \label{eq:c2alpha-wp}\\
    \hat{c}_{1\alpha}(t) &= -\frac{1}{2}D(g,2\,\alpha)^* e^{i\phi} c_g(t)F_{2\omega}(t)  \sin \left(\frac{1}{2}\int_{t}^\infty \Omega (t^\prime) dt^\prime\right) , \label{eq:c1alpha-wp}\\
    \hat{c}_{2\beta}(t) &= -\frac{1}{2}D(g,1\,\beta)^*  c_g(t)F_\omega(t)  \sin \left(\frac{1}{2}\int_{t}^\infty \Omega (t^\prime) dt^\prime\right) , \label{eq:c2beta-wp}\\
    \hat{c}_{1\beta}(t) &= -\frac{i}{2}D(g,1\,\beta)^*  c_g(t)F_\omega(t)  \cos \left(\frac{1}{2}\int_{t}^\infty \Omega (t^\prime) dt^\prime\right) . \label{eq:c1beta-wp}
\end{align}
Indeed, the oscillations seen in Fig.~\ref{fig:energy-dependence-real} is due to the interference between the wavepacket $\hat{c}_{1\alpha}$ and $\hat{c}_{1\beta}$ and between $\hat{c}_{2\alpha}$ and $\hat{c}_{2\beta}$, rather than optical interference.
By use of Parseval's theorem, we find,
\begin{align}
   & \int_{-\infty}^\infty \left(c_{1\alpha}(\Delta\epsilon,\infty) \left[c_{1\beta}(\Delta\epsilon,\infty)\right]^*+c_{2\alpha}(\Delta\epsilon,\infty) \left[c_{2\beta}(\Delta\epsilon,\infty)\right]^*\right) d\Delta\epsilon \nonumber \\
   & = \int_{-\infty}^\infty \left(\hat{c}_{1\alpha}(t) \left[\hat{c}_{1\beta}(t)\right]^*+\hat{c}_{2\alpha}(t) \left[\hat{c}_{2\beta}(t)\right]^*\right) dt = 0,
   \label{eq:energy-integration}
\end{align}
therefore $\int_{-\infty}^\infty (|c_{1\alpha}|^2+|c_{2\alpha}|^2+|c_{1\beta}|^2+|c_{2\beta}|^2)I_{\rm coh}(\Delta\epsilon) d\Delta\epsilon=0$.
This result indicates that energy-integrated PAD measurement cannot detect the control of coherence and entanglement by the Rabi oscillations and that energy-resolved measurement is required.

\subsection{Numerical examples for Ne}

Let us now turn to specific numerical calculations for a Ne atom subject to two-color XUV pulses. We consider that the fundamental and second-harmonic pulses have Gaussian intensity profiles with an FWHM duration of 40 and 30 fs, respectively, and a peak intensity of $8.9\times 10^{11}\,{\rm W/cm}^2$ and $2.0\times 10^{13}\,{\rm W/cm}^2$, respectively.
We use the following values for dipole transition matrix elements (in a.u.) \cite{Becker1996Springer}, $D(g,1\,\beta) = 0.7566, D(g,2\,\alpha) = 0.1371$, and $D(1,2)=0.2984$, leading to $\int_{-\infty}^\infty \Omega (t) dt = 3.74$ or $\xi=0.596$.

The photoelectron angular distribution $I_e(\theta,\Delta\epsilon$), an observable often used in coherent control experiments \cite{Prince2016NPhoton,Iablonskyi2017PRL,DiFraia2019PRL,You2020PRX}, is given by,
\begin{align}
I_e(\theta,\Delta\epsilon)&=\left|c_{1\alpha}(\Delta\epsilon,\infty)e^{i\eta_p}Y_{10}(\theta)-c_{1\beta}(\Delta\epsilon,\infty)\left[a_de^{i\eta_d}Y_{20}(\theta)+a_se^{i\eta_s}Y_{00}(\theta)\right]\right|^2 \nonumber\\
    &+ \left|c_{2\alpha}(\Delta\epsilon,\infty)e^{i\eta_p}Y_{10}(\theta)-c_{2\beta}(\Delta\epsilon,\infty)\left[a_de^{i\eta_d}Y_{20}(\theta)+a_se^{i\eta_s}Y_{00}(\theta)\right]\right|^2 \nonumber\\
    &+|B(\Delta\epsilon,\infty)|^2(|Y_{2,-1}(\theta,\varphi)|^2+|Y_{2,1}(\theta,\varphi)|^2),
    \label{eq:PAD}
\end{align}
where $\eta_s=0.8335,\eta_p=-0.6734,\eta_d=0.0145$ denote the scattering phase shifts, $a_s=0.5499, a_d=0.8352$ the branching amplitudes to the $s$ and $d$ partial waves, respectively, obtained by multiconfiguration Hartree-Fock calculations \cite{You2020PRX,Gryzlova2018PRA,Froese-Fischer1997}, and $B(\Delta\epsilon,t)$ the amplitude of non-interfering photoionization pathways from $2p\,(m=\pm 1)$ by the $\omega$ pulse, whose corresponding dipole coupling is 0.5473.
$I_e(\theta,\Delta\epsilon)$ can also be expressed as,
\begin{equation}
    I_e(\theta,\Delta\epsilon)=\frac{|c_{1\alpha}|^2+|c_{1\beta}|^2+|c_{2\alpha}|^2+|c_{2\beta}|^2+2|B|^2}{4\pi}\left[1+\sum_{l=1}^4\beta_l P_l(\cos\theta)\right],
    \label{eq:PAD2}
\end{equation}
where $P_l(\cos\theta)$ are the Legendre polynomials and $\beta_l$ are the corresponding asymmetry parameters. After some algebra, we have,
\begin{align}
    \beta_1 &= -\frac{2\sqrt{15}}{5Z_0}\operatorname{Re}\left[(c_{1\alpha}c_{1\beta}^*+c_{2\alpha}c_{2\beta}^*)(2a_de^{i\eta_{pd}}+\sqrt{5}a_se^{i\eta_{ps}})e^{i\phi}\right] \label{eq:beta1} \\
    \beta_2 &= \frac{2}{7Z_0}\left[7(|c_{1\alpha}|^2+|c_{2\alpha}|^2)+(|c_{1\beta}|^2+|c_{2\beta}|^2 a_d (5a_d+7\sqrt{5}a_s\cos\eta_{ds})+5|B|^2 \right] \label{eq:beta2}\\
    \beta_3 &= -\frac{6\sqrt{15}}{5Z_0}a_d \operatorname{Re} \left[(c_{1\alpha}c_{1\beta}^*+c_{2\alpha}c_{2\beta}^*)e^{i(\eta_{pd}+\phi)}\right] \label{eq:beta3}\\
    \beta_4 &= \frac{6}{7Z_0}\left[3(|c_{1\beta}|^2+|c_{2\beta}|^2)a_d^2-4|B|^2\right],\label{eq:beta4}
\end{align}
where,
\begin{equation}
Z_0=|c_{1\alpha}|^2+|c_{1\beta}|^2+|c_{2\alpha}|^2+|c_{2\beta}|^2+2|B|^2.
\end{equation}
Then, the asymmetry of the electron emission, defined as the difference between the emission in one hemisphere ($0<\theta<\pi/2$) and the other ($\pi/2<\theta<\pi$), devided by the sum, is expressed as,
\begin{equation}
    \label{eq:asymmetry}
    A(\phi,\Delta\epsilon)=\frac{\beta_1(\phi,\Delta\epsilon)}{2}-\frac{\beta_3(\phi,\Delta\epsilon)}{8}.
\end{equation}
Both $\beta_1$ and $\beta_3$ contain 
\begin{equation}
    (c_{1\alpha}c_{1\beta}^*+c_{2\alpha}c_{2\beta}^*)e^{i\phi},
\end{equation}
which indicates that the asymmetry oscillates with, or is coherently controlled by, the $\omega$-$2\omega$ relative phase $\phi$, and the degree of coherence $I_{\rm coh}(\Delta\epsilon$) (Eq.~\ref{eq:degree-of-interference}) characterizes its (complex) amplitude.

\begin{figure}
    \centering
    \includegraphics[width=0.8\hsize]{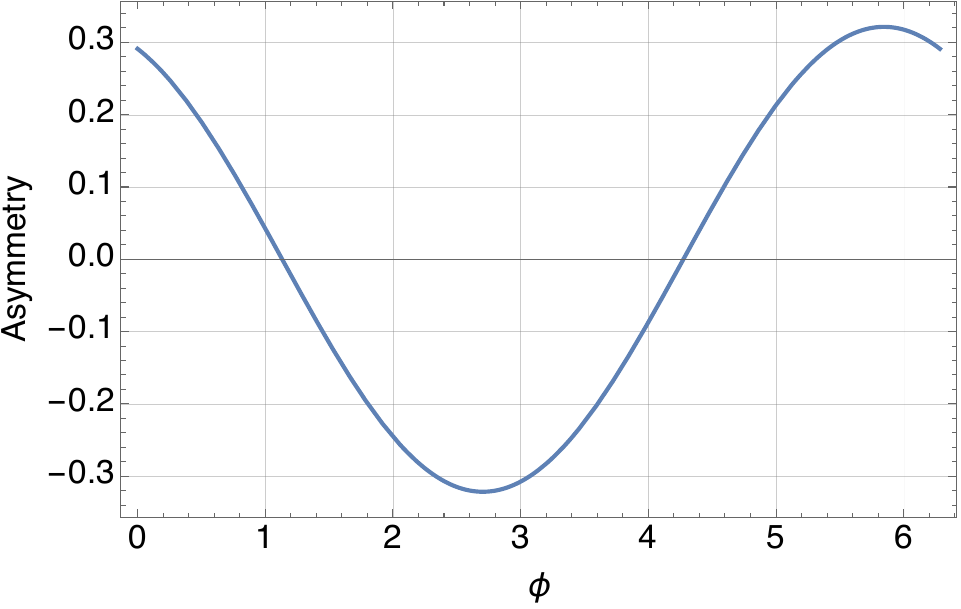}
    \caption{Asymmetry of the energy-resolved electron emission for the spectral center $\Delta\epsilon = 0$ and delay $\tau = -120\,{\rm fs}$.}
    \label{fig:Ne-asymmetry}
\end{figure}

\begin{figure}
    \centering
    \includegraphics[width=0.8\hsize]{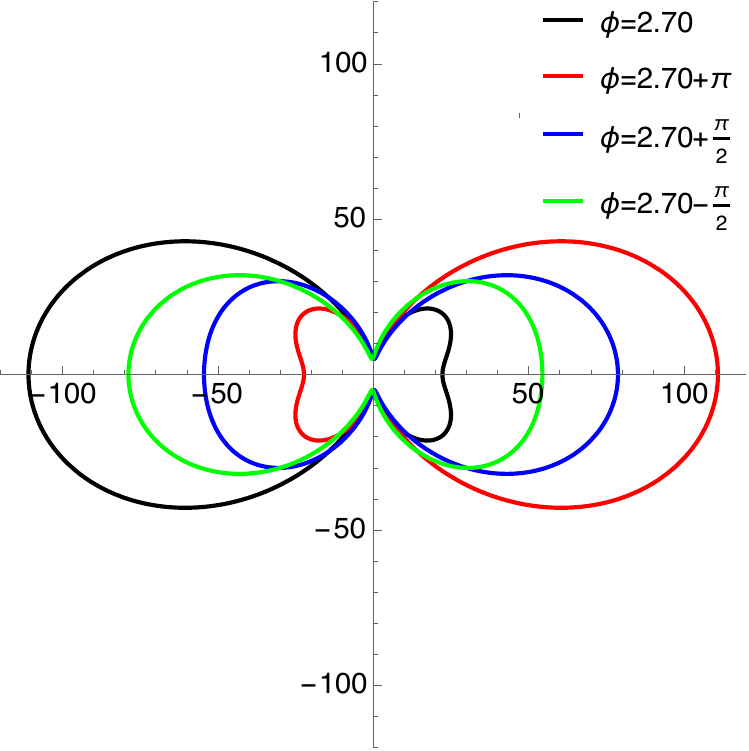}
    \caption{Energy-resolved photoelectron angular distribution at the spectral center $\Delta\epsilon = 0$ for several values of relative phase $\phi$.}
    \label{fig:Ne-PAD-center}
\end{figure}

Figure \ref{fig:Ne-asymmetry} shows $A(\phi,\Delta\epsilon)$ vs. $\phi$ at the spectral center $\Delta\epsilon = 0$ for the delay $\tau = -120\,{\rm fs}$ (the $2\omega$ pulse precedes the $\omega$ one by 120 fs).
The ground-state population after the pulses $c_g(\infty)^2$ is 0.539, where we have also taken ionization of Ne $2p$ by the $2\omega$ pulse into account.
As we expected, the asymmetry oscillates with the relative phase and reaches maximum ($\pm 0.322$) at $\phi=2.70$ and 5.85.
We show how the PAD changes with $\phi$ in Fig.~\ref{fig:Ne-PAD-center}. It is interesting to see that the PAD itself is not symmetric even for $\phi=2.70\pm \frac{\pi}{2}$, for which $A(\phi,\Delta\epsilon)=0$.

\begin{figure}
    \centering
    \includegraphics[width=0.8\hsize]{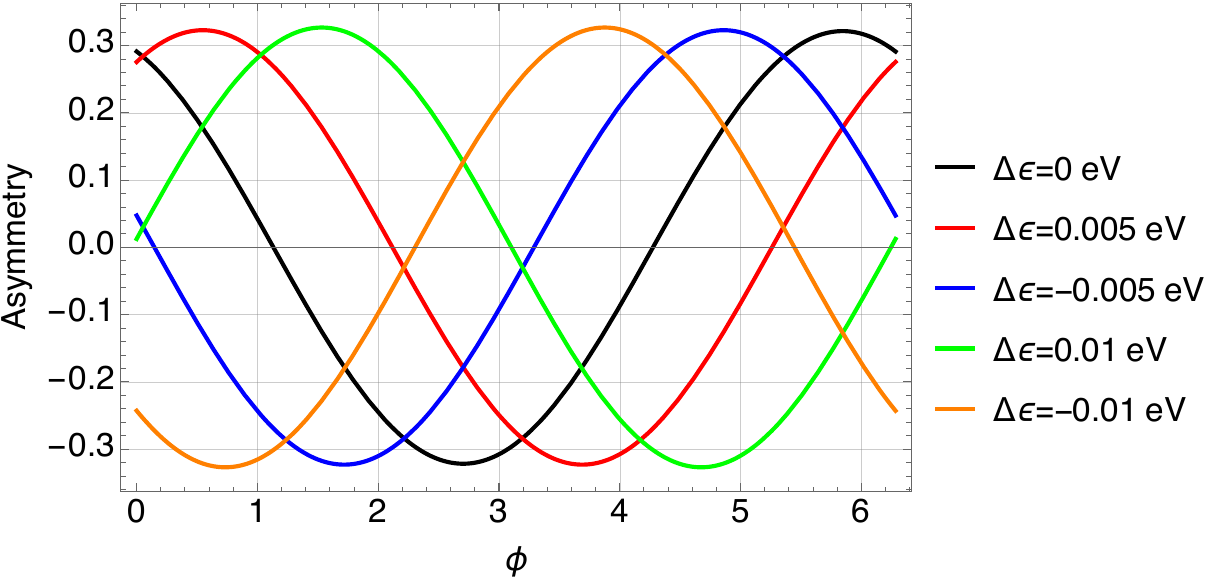}
    % Ne2s2p20210504-12（遅延120 fs）.nb
    \caption{Asymmetry of the energy-resolved electron emission for several values of $\Delta\epsilon$ and delay $\tau = -120\,{\rm fs}$.}
    \label{fig:Ne-asymmetry-energy-dependence}
\end{figure}

\begin{figure}
    \centering
    \includegraphics[width=0.8\hsize]{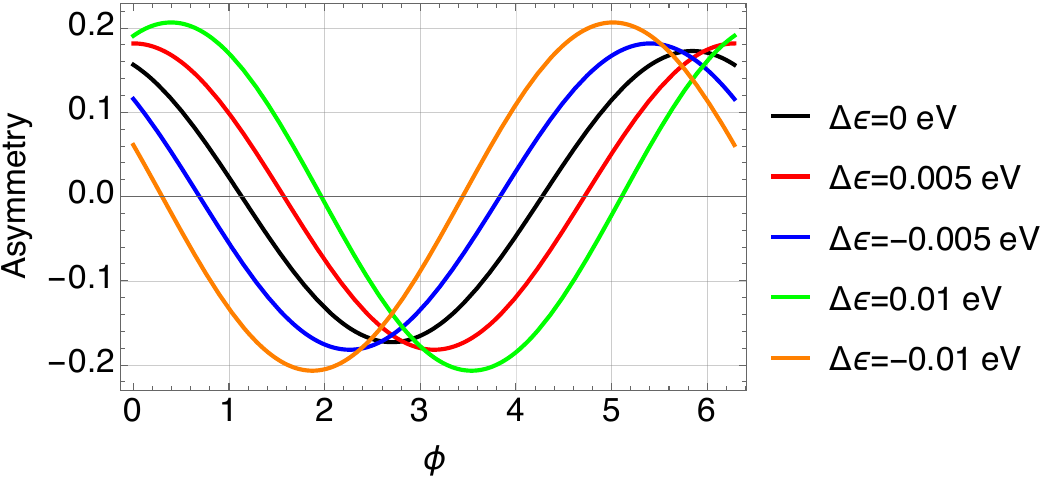}
    % Ne2s2p20210504-12（遅延20 fs）.nb
    \caption{Asymmetry of the energy-resolved electron emission for several values of $\Delta\epsilon$ and delay $\tau = -20\,{\rm fs}$.}
    \label{fig:Ne-asymmetry-energy-dependence-20fs}
\end{figure}

\begin{figure}
    \centering
    \includegraphics[width=0.8\hsize]{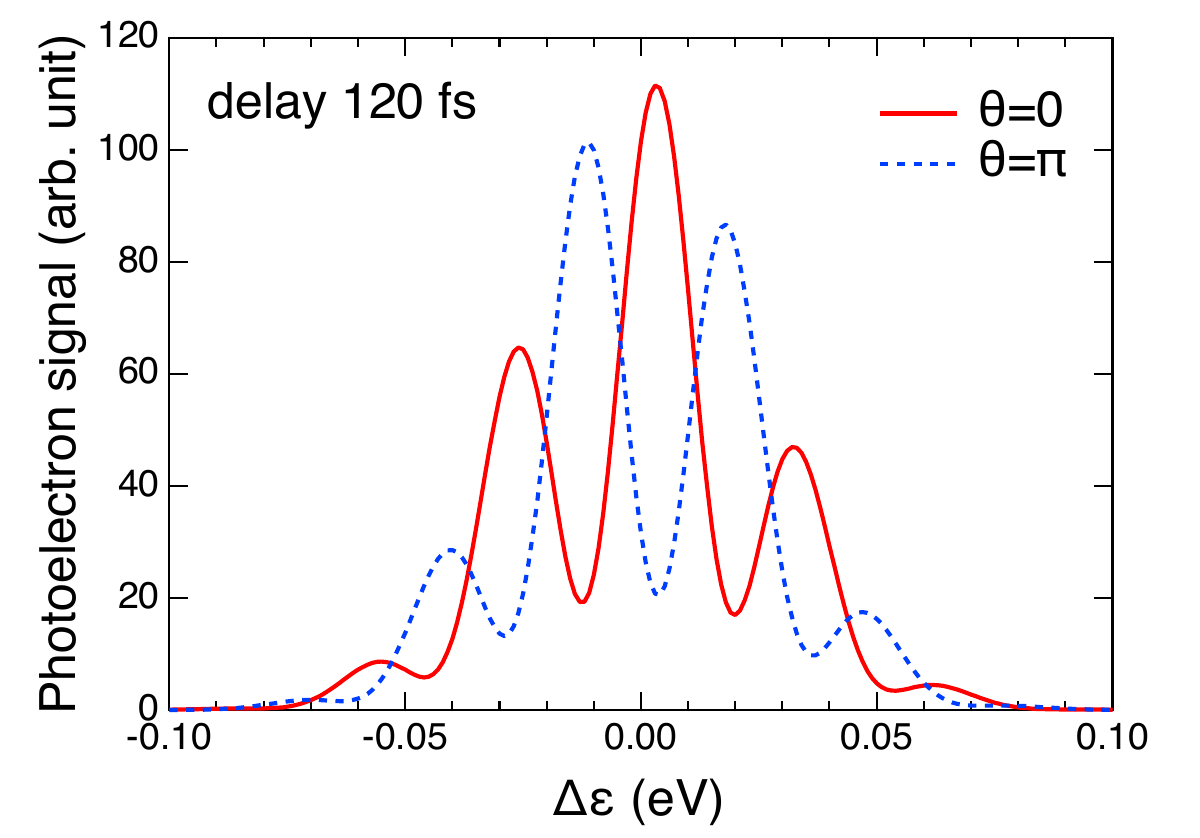}
    \caption{Photoelectron signal at $\theta=0$ and $\pi$ vs. $\Delta\epsilon$ for $\tau=-120$ fs.}
    \label{fig:signal120fs}
\end{figure}

\begin{figure}
    \centering
    \includegraphics[width=0.8\hsize]{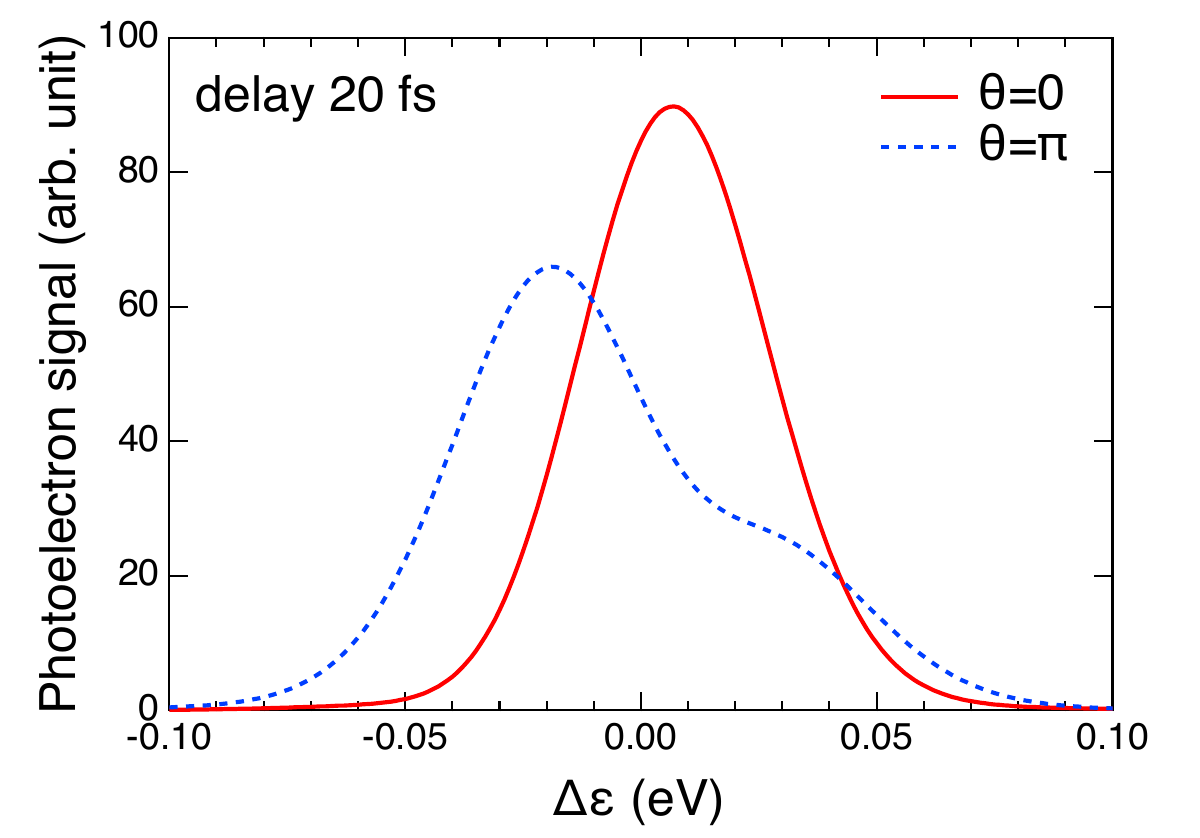}
    \caption{Same as Fig.~\ref{fig:signal120fs} for $\tau=-20$ fs.}
    \label{fig:signal020fs}
\end{figure}

Figures \ref{fig:Ne-asymmetry-energy-dependence} and \ref{fig:Ne-asymmetry-energy-dependence-20fs} exhibit how the asymmetry oscillation changes with $\Delta\epsilon$ for $\tau=-120$ and $-20\,{\rm fs}$, respectively.
Note that the FWHM spectral width of the $\omega$ and $2\omega$ pulses is 0.046 and 0.061 eV, respectively.
As we can expect from Fig.~\ref{fig:energy-dependence-real}, the oscillation curve shifts with $\Delta\epsilon$, which will cancel out after energy integration (Eq.~\ref{eq:energy-integration}).
The shift is smaller for smaller delay (Fig.~\ref{fig:Ne-asymmetry-energy-dependence-20fs}).
The oscillation amplitude is symmetric with respect to $\Delta\epsilon=0$, reflecting the symmetry in Fig.~\ref{fig:energy-dependence-magnitude}.
The variation of the asymmetry of the electron emission with $\Delta\epsilon$ can be clearly seen in the calculated photoelectron signal vs. $\Delta\epsilon$ for directions $\theta=0$ and $\phi$ (Figs.~\ref{fig:signal120fs} and \ref{fig:signal020fs}).

\subsubsection{Effect of focal volume averaging}

\begin{figure}
    \centering
    \includegraphics[width=0.8\hsize]{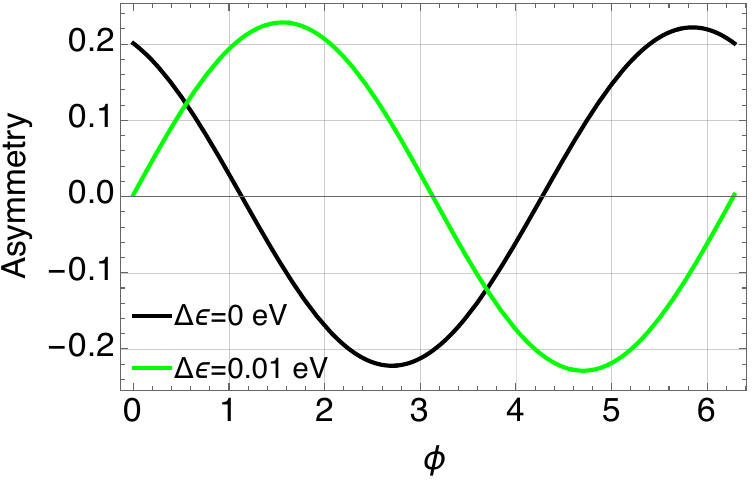}
    % Ne2s2p20210504-12（遅延120 fs）FVA.nb
    \caption{Asymmetry of the energy-resolved electron emission for $\Delta\epsilon=0$ and 0.01 eV and delay $\tau = -120\,{\rm fs}$. with focal volume averaging.}
    \label{fig:Ne-asymmetry-energy-dependence-120fs-fva}
\end{figure}

\begin{figure}
    \centering
    \includegraphics[width=0.8\hsize]{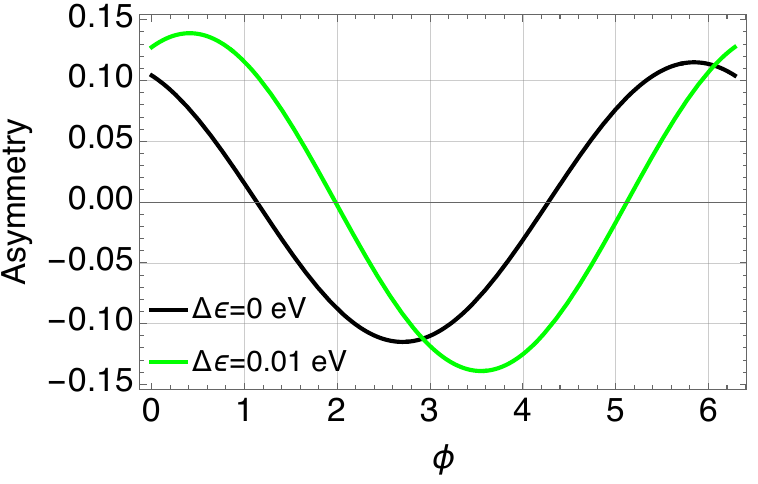}
    % Ne2s2p20210504-12（遅延20 fs）FVA.nb
    \caption{Asymmetry of the energy-resolved electron emission for $\Delta\epsilon=0$ and 0.01 eV and delay $\tau = -20\,{\rm fs}$. with focal volume averaging.}
    \label{fig:Ne-asymmetry-energy-dependence-20fs-fva}
\end{figure}

Let us briefly discuss how focal volume averaging affects the PAD asymmetry shown above. 
We assume that the fundamental and second-harmonic pulses have a common Gaussian beam profile with the same peak intensities on the axis as those used above, i.e., $8.9\times 10^{11}\,{\rm W/cm}^2$ and $2.0\times 10^{13}\,{\rm W/cm}^2$, respectively.
Then, we average the calculated PAD over the intensity distribution.
The PAD asymmetry vs.~$\phi$ is shown in Figs.~\ref{fig:Ne-asymmetry-energy-dependence-120fs-fva} and \ref{fig:Ne-asymmetry-energy-dependence-20fs-fva} for $\tau=-120$ and $-20\,{\rm fs}$, respectively.
Comparing them with Figs.~\ref{fig:Ne-asymmetry-energy-dependence} and \ref{fig:Ne-asymmetry-energy-dependence-20fs}, respectively, we see that the PAD can still be coherently controlled after focal volume averaging.
Although it may come as a surprise, this result can be understood for $\Delta\epsilon=0$ by from Fig.~\ref{fig:xi-dependence} as follows.

The degree of coherence $I_{\rm coh}$ is pure imaginary at $\Delta\epsilon=0$ and $\phi=0$, and its imaginary part is plotted in Fig.~\ref{fig:xi-dependence}. As mentioned above, the present calculation conditions correspond to $\xi=0.596$, i.e., half Rabi cycle. We can see from the figure that $\operatorname{Im}I_{\rm coh}(\Delta\epsilon=0)|_{\phi=0}$ is monotonic in the interval $\xi\in [0,0.596]$ and does not change sign. Thus, combined with Eq.~\ref{eq:Icoh-phi-dependence}, the contributions from all the intensities across the beam profile add up constructively and in phase, as seen in Figs.~\ref{fig:Ne-asymmetry-energy-dependence-120fs-fva} and \ref{fig:Ne-asymmetry-energy-dependence-20fs-fva}.

\section{Conclusions}
\label{sec:conclusions}

We have theoretically analyzed the photoionization of one shell of an atom (e.g., Ne $2p$) by a fundamental frequency and a second shell (e.g., Ne $2s$) by the coherent second harmonic, for the case that the two processes emit electrons of the same energy, and for linear, parallel polarization. Under these conditions, the fundamental energy is necessarily equal to the energy difference of the two hole states, and therefore induces Rabi coupling. 
In the absence of this coupling the photoelectron wave packets from the two shells would not interfere even though they have the same energy, since they are associated with different ionic states.
The Rabi coupling recovers coherence, which can be observed as asymmetry in the energy-resolved photoelectron angular distribution, controlled through the $\omega$-$2\omega$ relative phase $\phi$. Interference is observed as an asymmetry in the PAD and varies as a function of the phase difference between the two pulses. 
For the special case of identical pulse shapes, zero delay between the pulses, and photoelectron energy equal to the central energy of the photoelectron wave packet, the asymmetry vanishes. 
When there is a delay between the pulses, stronger asymmetry is induced, and we have studied numerically the effects of Rabi coupling strength, pulse shape and photoelectron energy. 
We find that energy-resolved measurement is necessary to observe the asymmetry, because integration over the photoelectron energy causes the asymmetry to vanish. The results have been derived for a generic system, and also for the particular case of ionization of the $2s$ and $2p$ shells of Ne, with the inclusion of experimental parameters. 
Our model predicts the phase dependence of the PAD will be observable even for long delays between the pulses, corresponding to wave packets which are far from the ion core when it is modified by the second delayed pulse. 
This is a manifestation of photoelectron - ion core entanglement in the ionization process.

It will be useful to briefly mention experimental feasibility.
To perform the experiment described here on Ne, it is clear that both angular and energetic resolution are required, as well as the possibility to produce delayed, phase-coherent pulses. The last requirement is fulfilled at the FERMI FEL, where it has been demonstrated that two coherent pulses with delay similar to that used in the example above can be produced \cite{Perosa2023PRL}, or with longer delays, at the cost of some phase jitter. In that work, pulses of opposite circular polarization were produced, but the same schemes can be used to produce different harmonics with the same polarization.
The energy resolution required is equal to or better than the fringe spacing ($\sim h/\tau$), which is about 8 meV for well-separated pulses ($\tau=50$ fs). Regarding detectors, the commonly used velocity map imaging spectrometer is unsuitable, as it generally has poor energy resolution. Magnetic bottle spectrometers have good energy resolution, especially when retardation is used, but generally insufficient angular resolution. However, it should be noted that the instrument can be set to accept one hemisphere of the PAD, and this angular resolution was sufficient to characterize attosecond pulse trains \cite{Maroju2020}, where the key observable was asymmetry with respect to the direction of the electric vector.  Thus, this angular resolution might be sufficient. The best spectrometer for this experiment appears to be an array of time-of-flight detectors, which have high angular and energy resolution \cite{Viefhaus2013NIMA}, and indeed, an instrument of this class is available as a diagnostic at European X-Ray Free Electron Laser.

\begin{acknowledgement}

This research was supported in part by a Grant-in-Aid for Scientific Research (Grant No. JP19H00869 and JP20H05670) from the Ministry of Education, Culture, Sports, Science and Technology (MEXT) of Japan and by JST CREST (Grant No. JPMJCR15N1).
The authors thank Carlo Callegari for discussions on the machine parameters of FERMI and Elena V. Gryzlova for providing us with the calculated values of $\eta_s,\eta_p,\eta_d,a_s$, and $a_d$.

% Please use ``The authors thank \ldots'' rather than ``The
% authors would like to thank \ldots''.

% The author thanks Mats Dahlgren for version one of \textsf{achemso},
% and Donald Arseneau for the code taken from \textsf{cite} to move
% citations after punctuation. Many users have provided feedback on the
% class, which is reflected in all of the different demonstrations
% shown in this document.

\end{acknowledgement}

% \begin{suppinfo}

% This will usually read something like: ``Experimental procedures and
% characterization data for all new compounds. The class will
% automatically add a sentence pointing to the information on-line:

% \end{suppinfo}

%\bibliography{achemso-demo}
\bibliography{PhysRep}

\end{document}